\documentclass[twocolumn,aps,prb,showpacs]{revtex4-1}
\usepackage{graphicx}
\usepackage{dcolumn}
\usepackage{bm}
\begin{document}
\title{Topological electronic structure and Weyl semimetal in the TlBiSe$_2$ class of semiconductors}
\author{Bahadur Singh$^1$, Ashutosh Sharma$^1$, H. Lin$^2$, M.Z. Hasan$^3$$^,$$^4$, R. Prasad$^1$ and A. Bansil$^2$}
\affiliation{$^1$Department of Physics, Indian Institute of Technology Kanpur, Kanpur 208016, India \\
             $^2$Department of Physics, Northeastern University, Boston, Massachusetts 02115, USA \\
           $^3$Joseph Henry Laboratory, Department of Physics, Princeton University, Princeton, New Jersey 08544, USA\\
             $^4$Princeton Center for Complex Materials, Princeton University, Princeton, New Jersey 08544, USA }

\begin{abstract}
We present an analysis of bulk and surface electronic structures of thallium based ternary III-V-VI$_2$ series of compounds TlMQ$_2$, where M=Bi or Sb and Q=S, Se or Te,
using the \emph{ab initio} density functional theory framework. Based on parity analysis and (111) surface electronic structure, we predict TlSbSe$_2$, TlSbTe$_2$, TlBiSe$_2$ and 
TlBiTe$_2$ to be non-trivial topological insulators with a single Dirac cone at the $\Gamma$-point, and TlSbS$_2$ and TlBiS$_2$ to be trivial band insulators. Our predicted topological
phases agree well with available angle-resolved photoemission spectroscopy (ARPES) measurements, in particular the topological phase changes between TlBiSe$_2$ and TlBiS$_2$. Moreover, 
we propose that Weyl semimetal can be realized at the topological critical point in TlBi(S$_{1-x}$Se$_x$)$_2$ and TlBi(S$_{1-x}$Te$_x$)$_2$ alloys by breaking the inversion symmetry in the
layer by layer growth in the order of Tl-Se(Te)-Bi-S, yielding six Dirac cones centered along the $\Gamma-L$ directions in the bulk band structure.  
\end{abstract}

 
\maketitle

\section{introduction}

Topological insulators are a new class of materials, which have attracted intense interest in the last few years\cite{xl,hasan,moore} due to their exotic properties. These
materials support an odd number of surface state bands with linear dispersion in the bulk energy gap, which can be viewed as a sea of massless Dirac Fermions. The conducting surface states in 
topological insulators are protected by time-reversal symmetry and are immune to scattering by non-magnetic impurities, thus opening new avenues for dissipationless transport. There also is a 
vigorous ongoing search for topological superconductors\cite{superconduct,superconduct1}, with the possibility of realizing Majorana fermions, which are their own 
antiparticles\cite{majorana} with potential application to quantum computing \cite{computing}. Topological insulators have also generated a considerable excitement due to the possibility of 
exploring the \textit{Higgs mechanism} and  realization of a Weyl semimetal in a condensed matter system. \cite{wan,burkov,halasz,suichi,weyl3,weyl7}

A topological phase was initially predicted in 2D HgTe/CdTe quantum wells \cite{bervenig} and subsequently verified experimentally \cite{konig}. Soon thereafter a 2D conducting surface state was
realized in the bulk band gap of three dimensional thermoelectrics \cite{zhang,hsieh,xia} Bi$_{1-x}$Sb$_x$, Bi$_2$Se$_3$, Bi$_2$Te$_3$ and Sb$_2$Te$_3$. However,
magnetotransport studies\cite{analytcis,cite,eto,butch} have shown that bulk transport dominates in these materials, motivating continued search for other 3D topological insulators
with single Dirac cone surface states residing in the bulk energy gap. Since then first-principles calculations have suggested a large variety of topologically interesting materials ranging from 
oxides\cite{shitade} to the Heusler family of compounds \cite{hlin,chadov,frenz}. Another class is thallium based ternary III-V-VI$_2$ chalcogenides, which were proposed 
theoretically \cite{lin,yan} and then  verified experimentally  \cite{kuroda,chen,sato}. These studies showed the existence of single Dirac cone type surface states
at the $\Gamma$-point in TlSbSe$_2$, TlSbTe$_2$, TlBiSe$_2$ and TlBiTe$_2$. Also, it has been found that $p$ doped TlBiTe$_2$ superconducts \cite{super}, where superconductivity is attributed to 
six leaf like bulk pockets in the Fermi surface and the surface state becomes superconducting\cite{chen}.

Interestingly, recent studies\cite{higgs,mass} of TlBi(S$_{1-x}$Se$_x$)$_2$ alloys show that a topological phase transition can be realized by modulating either spin-orbit coupling or 
the crystal structure, and that the surface Dirac fermion becomes massive at the quantum phase transition. This can be viewed as a condensed matter version of the \textit{Higgs mechanism} in
which a particle acquires a mass by spontaneous symmetry breaking. This system thus may provide a model system which connects condensed-matter physics to particle physics. It has also been 
proposed that a Weyl semimetal phase \cite{nielson1,nielson2} could be achieved at the phase transition between a topological and normal insulator if we explicitly break time reversal
symmetry\cite{wan,burkov} or inversion symmetry\cite{halasz,suichi}. In this new phase, valence and conduction bands touch at certain points, called Weyl points, where 
dispersion is linear. These Weyl points come in pairs with positive and negative helicities and are robust to perturbations in the bulk material.

Our motivation for undertaking the present study is to provide a comprehensive investigation of the bulk and surface electronic structures of the thallium based ternary III-V-VI$_2$ series of 
compounds TlMQ$_2$, where M=Bi or Sb and Q=S, Se or Te, within a uniform first principles framework. In particular, not only the nontrivial compounds are studied \cite{emereev_prb,banergee}
but the topologically trivial compounds TlBi(Sb)S$_2$ is included since the recent ARPES measurement observed the topological phase transition in the alloy TlBi(Se,S)$_2$ between nontrivial
TlBiSe$_2$ and trivial TlBiS$_2$.\cite{mass} We compare our theoretical predictions with the available ARPES results. Insight into the topological nature of these compounds is gained through
slab computations in which the thickness of the slab is varied. Finally, the possibility of realizing a Weyl semimetal phase through strain in TlBi(S$_{1-x}$Se$_x$)$_2$ and 
TlBi(S$_{1-x}$Te$_x$)$_2$ alloys is examined for the first time, and it is shown that, when the inversion symmetry of the system is broken, a Weyl phase is indeed possible at the topological 
critical point where the system undergoes a transition from a trivial to a non-trivial (topological) insulator. Bulk and surface electronic structures of many members of the six ordered 
Tl-compounds considered in this work have been discussed in several earlier papers in the literature. \cite{lin}$^,$\cite{yan}$^,$\cite{emereev_prb}$^,$\cite{banergee}. 

The organization this article is as follow. Section II gives details of the bulk and surface computations. In section III, we explain the bulk crystal structures, electronic structure and the 
parity analysis used to infer the topological nature of various compounds. Section IV discusses the slab structure used, relaxation and size dependent effects. In section V, we explain the
topological phase transition and realization of the Weyl semimetal phase in TlBi(S$_{1-x}$Se$_x$)$_2$ and TlBi(S$_{1-x}$Te$_x$)$_2$ alloys. Section VI summarizes conclusions of this study.

\section{Computational Details}
Electronic structure calculations were carried out within the framework of the density functional theory \cite{kohan} using VASP\cite{vasp}($Vienna$ $ab$ $initio$ $Simulation$ $Package$),
with projected augmented wave basis\cite{paw}. Exchange-correlation effects were treated using a generalized gradient approximation \cite{pbe} and the 
spin-orbit coupling (SOC) effects are included as implemented in the VASP package. For structure optimization a plane wave cut-off energy of 350 eV 
and a $\Gamma$-centered 8$\times$8$\times$8 k-mesh with conjugate gradient algorithm(CGA)\cite{cga}was used. Lattice parameters and ionic positions were adjusted until all components of 
Hellman-Feynman force on each ion were less than 0.001 eV/\r{A}. Our surface electronic structure calculations are based on a slab geometry using bulk relaxed parameters with a net vacuum of
15\r{A}, a plane wave cut-off energy of 350 eV, and a $\Gamma$-centered 9$\times$9$\times$1 k-mesh. Since ionic relaxations \cite{emereev_prb,banergee} are important, ionic positions 
in all slabs were optimized until the $z$-component of the Hellman-Feynman forces was less than 0.005 eV/\r{A}. Topological phase transition in TlBi(S$_{1-x}$Se$_x$)$_2$ and
TlBi(S$_{1-x}$Te$_x$)$_2$ systems was investigated by taking the critical concentration of sulfur to be x=0.5\cite{mass}.

\section{Results and Discussion} 

\subsection{Bulk Crystal Structure}

Thallium based ternary chalcogenides III-V-VI$_2$ share a rhombohedral crystal structure (space group D$^5_{3d}$ (R$\bar{3}$ m)) with four atoms per unit cell which occupy the Wyckoff positions
 3a, 3b, and 6c \cite{white}. We illustrate the crystal structure with the example of TlBiTe$_2$, which can be viewed as a distorted NaCl structure with four atoms in the primitive unit
cell \cite{white,handbook,hoang} and a sequence of hexagonal close packed layers in the order Tl-Te-Bi-Te (see Fig.1).
\begin{figure}[ht!] 
\includegraphics[width=0.5\textwidth]{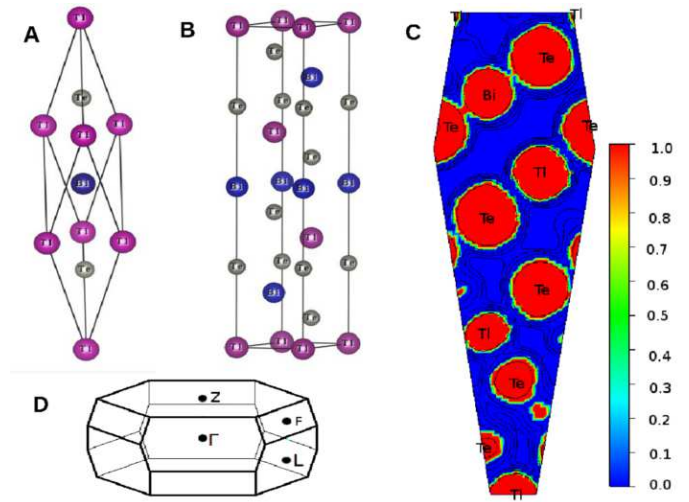}
\caption{Primitive (A) and conventional (B) hexagonal crystal structure for TlBiTe$_2$. The conventional (hexagonal) unit cell has three times more atoms than primitive
cell.(C) ELF plot of TlBiTe$_2$ (see text). 
(D) Brillouin zone for the primitive unit cell with four time reversal invariant points $\Gamma$, F, Z and L.} 
\end{figure}
The conventional unit cell is hexagonal and contains 12 atoms.  The hexagonal lattice
constants $a_H$ and $c_H$ can be computed from the rhombohedral parameters via the relation
\begin{equation}
a_{H}=2a_{R}\sin\frac{\alpha}{2}, \quad   c_{H}=a_{R}\sqrt{3+6\cos{\alpha}}
\end{equation}
where a$_R$ is the rhombohedral lattice constant and $\alpha$ is the rhombohedral angle. Notably, the present compounds form a layered structure similar to  binary topological insulators 
A$_2$B$_3$ (A=Bi, Sb and B=Se, Te)\cite{zhang} where atoms are arranged in a sequence of five atomic layers or a quintuple repeating pattern. The bonding within the quintuple of layers is 
ionic-covalent type while between one quintuple group and the next is van-der Waals type \cite{mishra_bi2se3_bond}. In order to gain insight into the nature of bonding in the thallium based
chalcogenides we have performed an ELF (Electron Localization Function)\cite{elf} study, which shows that there is a negligible value of the ELF between the thallium and tellurium layers, 
indicating predominantly ionic type bonding between these layers. In contrast, the value of ELF between the bismuth and tellurium layers is significant implying ionic-covalent type bonding.
The optimized lattice and internal parameters $u$ for the six Tl-compounds considered are given in Table 1. 
\begin{table}[h!]
\centering
\caption{Optimized lattice constant $a_{R}$, angle $\alpha$ and internal parameter $u$ for six thallium based rhombohedral compounds.}
\begin{ruledtabular}
\begin{tabular}{  c c  c  c }
 Compound   &   $a_{R}$      &$\alpha$     & $u$\\
            &   (\r{A})      &(degrees)         \\
\hline 

TlSbSe$_2$  &      7.955     & 30.869       & 0.2374\\
TlSbTe$_2$  &      8.421     & 31.367       & 0.2386\\
TlSbS$_2$   &      7.718     & 30.568       & 0.2352\\
TlBiSe$_2$  &      7.989     & 31.368       & 0.2391\\
TlBiTe$_2$  &      8.367     & 31.847       & 0.2413 \\ 
TlBiS$_2$   &      7.979     & 29.764       & 0.2368\\

\end{tabular}
\end{ruledtabular}
\end{table}

\subsection{Bulk Band Structures}

The bulk band structures of TlSbQ$_2$ {Q=S, Se, and Te}, shown in Fig 2, indicate these three compounds to be narrow gap semiconductors. TlSbSe$_2$ is seen to be a direct band gap semiconductor 
with the valence band maximum (VBM) and conduction band minimum (CBM) lying along $\Gamma-L$. Here the spin-orbit coupling plays an important role as it induces a band inversion at $\Gamma$ 
suggesting a possible nontrivial topological phase, a point to which we return below. TlSbTe$_2$ and TlSbS$_2$ are also seen from Figs. 2(B)-(C) to be direct band gap semiconductors with VBM 
and CBM at the $\Gamma$-point. Since Te is heavier than Se, there is a large spin-orbit (SOC) effect in TlSbTe$_2$.
But S atom is lighter than Se and Te so that SOC effects are smaller in TlSbS$_2$. 
\begin{figure}[h!]
\includegraphics[width=0.5\textwidth]{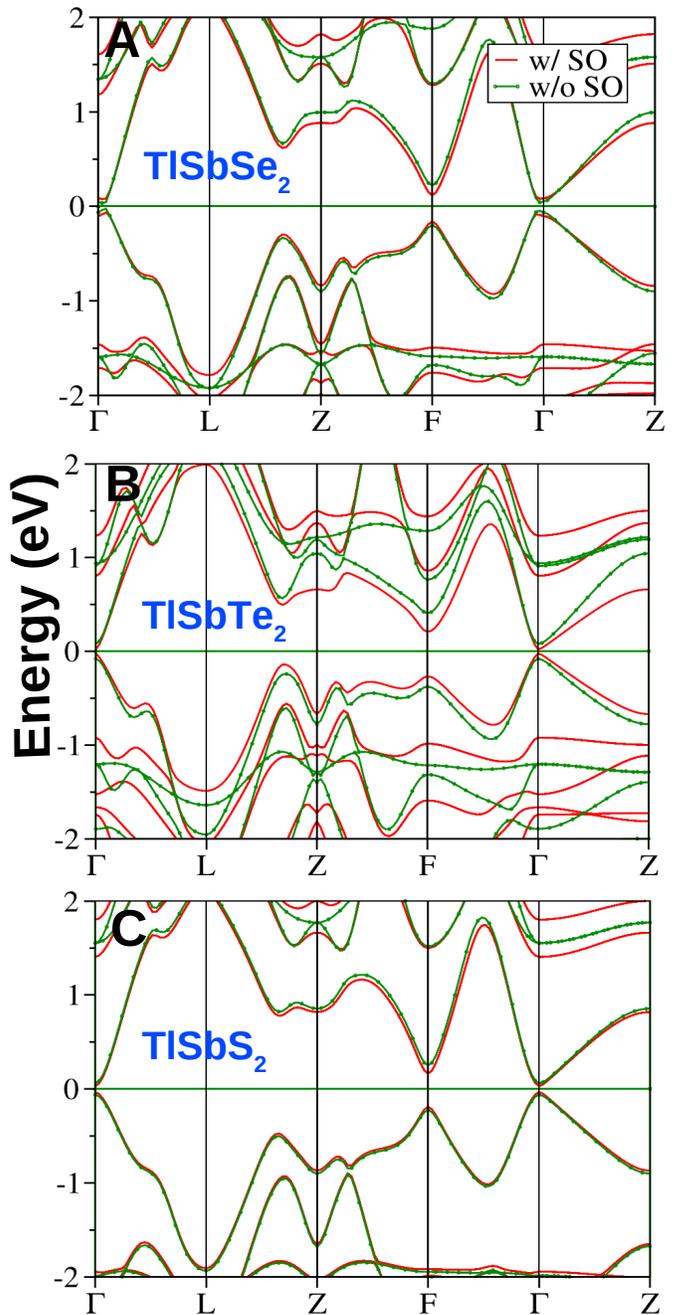}
\caption{Bulk band structures of rhombohedral (A) TlSbSe$_2$ (B) TlSbTe$_2$ (C) TlSbS$_2$ along high symmetry 
lines with (red lines) and without (green lines) spin-orbit coupling.}
\end{figure}
\begin{figure}[h!]
\includegraphics[width=0.5\textwidth]{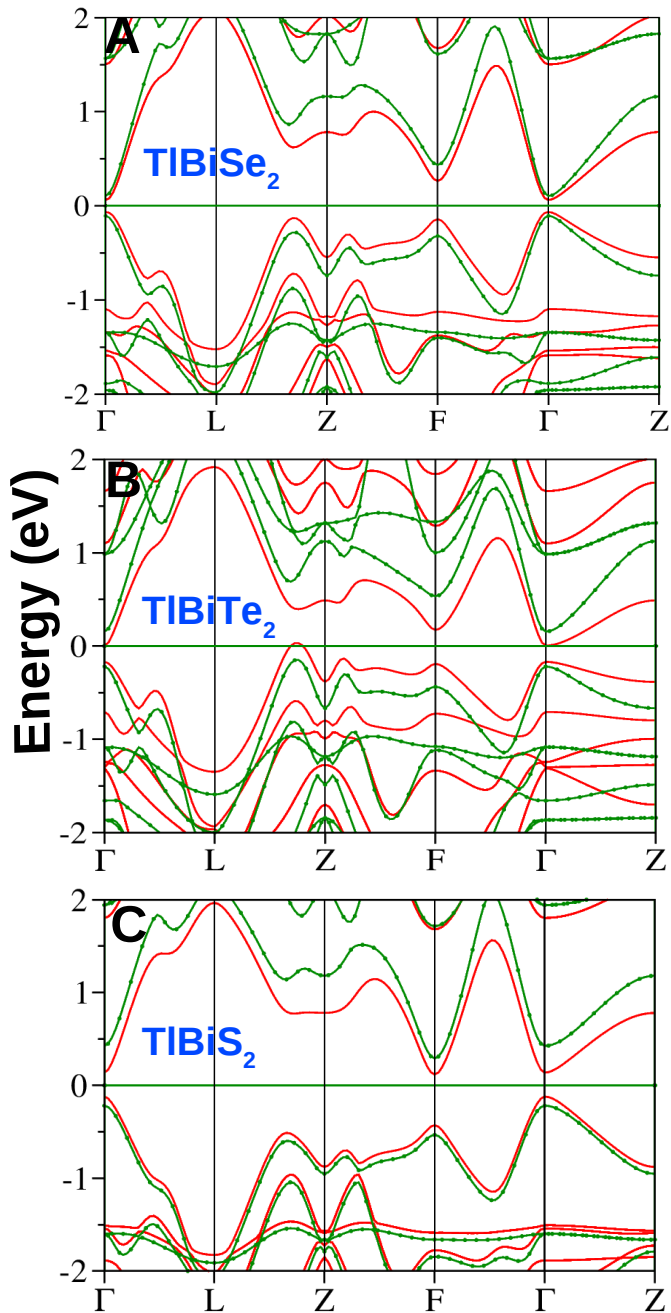}
\caption{Bulk band structures of rhombohedral (A) TlBiSe$_2$ (B) TlBiTe$_2$ (C) TlBiS$_2$ along high symmetry line with (red lines) and without (green lines) spin-orbit coupling.}
\end{figure}

The bulk band structures of TlBiQ$_2$ (Q=Se, Te, and S) are shown in Fig.3.  There is a large effect of spin-orbit coupling due to presence of the heavier Bi atom. 
Our band structures agree well with an earlier study\cite{hoang}, except in the case of TlBiTe$_2$, which was previously found to be \cite{hoang} an indirect gap semiconductor
with the VBM  lying along the $L$ $-$ $Z$ direction and the CBM at the $\Gamma$-point. However, for our relaxed structure (including spin-orbit coupling) the VBM and CVM still 
lie at the aforementioned $\vec{k}$-points, but the system is semi-metallic with a band gap of -10 meV. This is in accord with the corresponding experimental results\cite{chen}, 
which found the material to be semi-metallic with a band gap of -20 meV.

\subsection{Parity analysis}
Since all the investigated compounds possess inversion symmetry, a parity analysis\cite{kane} can be used to 
identify the Z$_2$ topological phases. There are eight time reversal invariant points in the rhombohedral Brillouin zone but only four points ($\Gamma$, F, L and Z; see Fig. 1 (D)) 
are inequivalent. Products of parity eigenvalues at these four momenta are given in Table II with and without spin-orbit coupling. 
\begin{table}[ht!]
\centering
\begin{ruledtabular}
\begin{tabular}{ c  c c  c c  c c c c   c  }

Compound  &  $\Gamma$  &$\times$1  & L  & $\times$3 & F  & $\times$3 & Z  & $\times$1  &  Z$_2$  \\
 & wso & so & wso & so & wso & so & wso & so & \\
\hline 
TlSbSe$_2$  & $+$ & $-$ & $+$ & $+$ & $+$ & $+$ & $+$  & $+$ & $1$ \\
TlSbTe$_2$  & $+$ & $-$ & $+$ & $+$ & $+$ & $+$ & $+$  & $+$ & $1$ \\
TlSbS$_2$   & $+$ & $+$ & $+$ & $+$ & $+$ & $+$ & $+$  & $+$ & $0$ \\ 
TlBiSe$_2$  & $+$ & $-$ & $+$ & $+$ & $+$ & $+$ & $+$  & $+$ & $1$ \\
TlBiTe$_2$  & $+$ & $-$ & $+$ & $+$ & $+$ & $+$ & $+$  & $+$ & $1$ \\
TlBiS$_2$   & $+$ & $+$ & $+$ & $+$ & $+$ & $+$ & $+$  & $+$ & $0$ \\
\end{tabular}
\end{ruledtabular}
\caption{Products of parity eigenvalues at the four inequivalent time-reversal invariant k-points in the six investigated compounds. 'wso' refers to without spin-orbit coupling 
and 'so' to with spin-orbit coupling.}
\end{table}

Table II, shows that product of parity eigenvalues in TlBiSe$_2$, TlBiTe$_2$, TlSbSe$_2$ and TlSbTe$_2$ changes at $\Gamma$ as spin-orbit
coupling is turned on, yielding a nontrivial topological invariant $Z_2=1$. The non-trivial topological character of these compounds is thus due to band inversion at 
the $\Gamma$-point \cite{lin}$^,$\cite{yan}$^,$\cite{emereev_prb}. On the other hand, there is no band-inversion for TlSbS$_2$ 
and TlBiS$_2$ at any of the time reversal invariant momenta, indicating that these compounds are topologically trivial.

\section{Surface Analysis}
\subsection{Slab Structure and Relaxation}
A hexagonal unit cell is used for all slab computations with atomic layers stacked in the z-direction. Since atoms 
along the z-direction are sequenced in the order Tl-Te(S, Se)-Bi(Sb)-Te(S, Se), there are four possible surface terminations, depending on which atom lies in the topmost layer. 
The bond length between Tl-Te is large (d=350 $pm$), whereas bond length between the Bi-Te is small (d=318 $pm$). 
Thus, out of the four possible surface terminations, we have used the one with Q (Se,Te and S) atom at the surface and Bi (Sb) under it in the second layer as this termination has 
a minimum number of dangling bonds \cite{lin}. Optimized bulk parameters were used to construct slabs of different thicknesses. A symmetric 39 atomic layer slab of TlBiTe$_2$
and the associated surface Brillouin zone is shown in Fig. 4. Surface relaxation is known to play an important role \cite{emereev_prb}$^,$\cite{banergee}, and accordingly, 
we have relaxed all atomic layers in the slabs used, although the relaxation effect is small as we move into the bulk.

\begin{figure}[h!]
\includegraphics[width=0.5\textwidth]{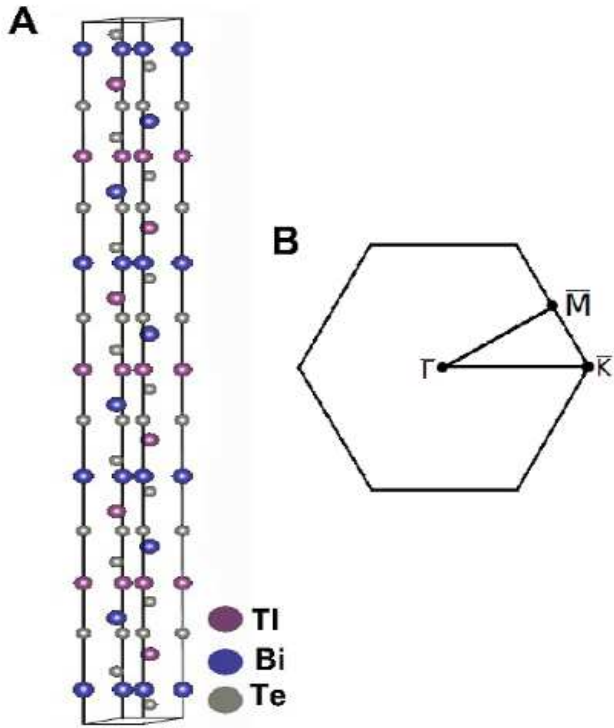}
\caption{(A) Structure of a 39 atomic layer slab of TlBiTe$_2$ stacked along the $z$-direction. (B) 2D Brillouin zone with three special k-points $\bar{\Gamma}$,$\bar{M}$ and $\bar{K}$.} 
\end{figure}

\subsection{TlSbQ$_2$ (Q=Se,Te and S)}

The Q-atom terminated surface electronic structures of the three compounds are 
shown in Fig. 5 for different slab thicknesses together with the associated projected bulk bands(blue colored region). Figs. 5(A) and (B) show the surface band structure of TlSbSe$_2$ 
for slabs of 35 (thickness $\approx$ 6.4 nm) and 47 (thickness $\approx$ 8.6 nm) layers, respectively. In the 35 layer slab there is a band gap of 50 meV at the $\bar{\Gamma}$-point. 
As the number of layers increases, the size of the gap decreases. For 47 layers, this gap is negligible and we get a clear Dirac-cone surface state in the bulk gap 
region. Moreover, around -0.8 eV we obtain a Rasbha-type, trivial spin split surface state in both slabs \cite{emereev_prb}. Fig. 5(C) shows results for a 47 layer TlSbTe$_2$ slab,
which are similar to those for TlSbSe$_2$. 
Finally, TlSbS$_2$ does not display any surface state, which confirms its topologically trivial nature.

\begin{figure}[h!] 
\includegraphics[width=0.5\textwidth]{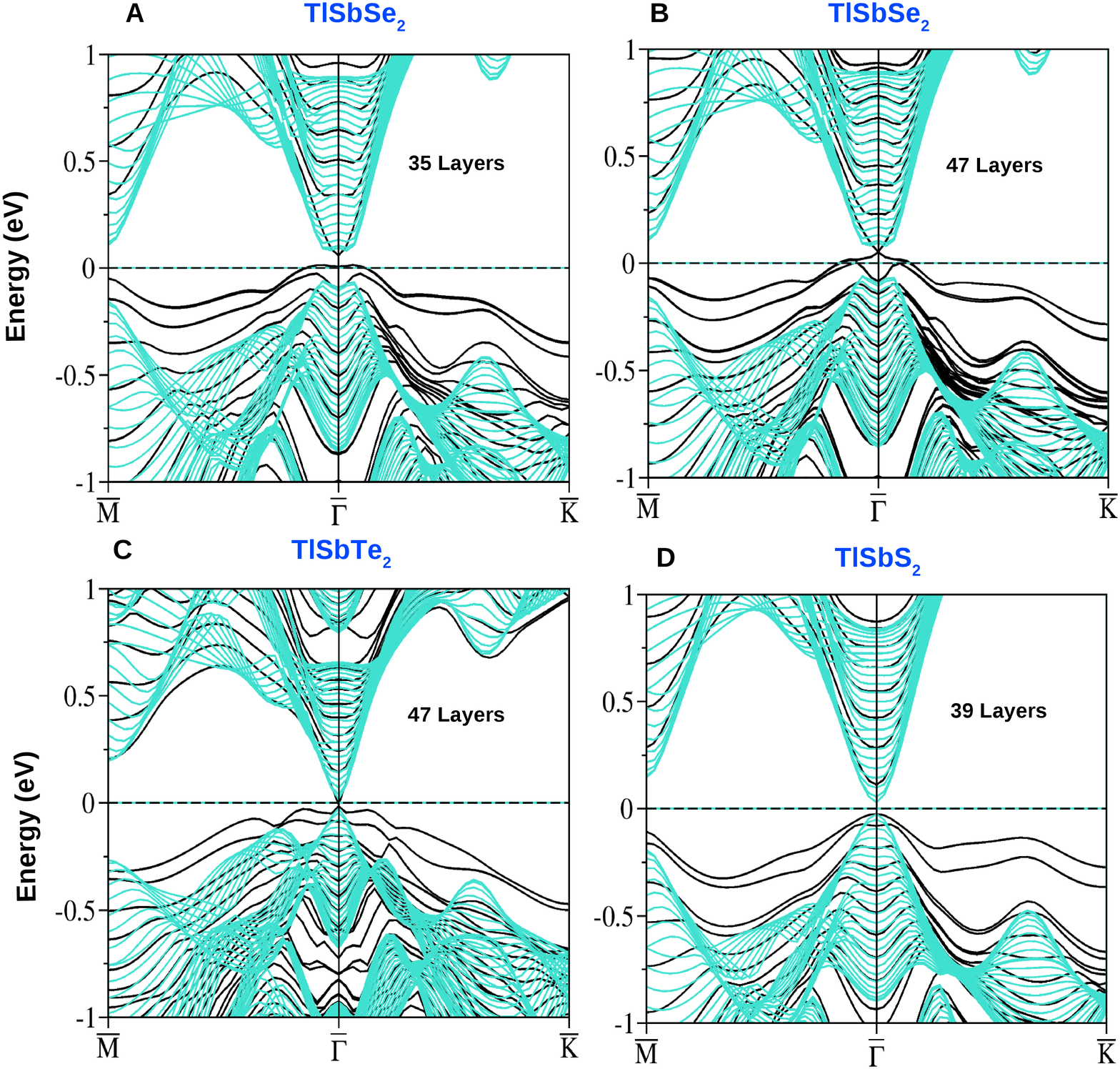}
\caption{Surface electronic structure of various slabs: 
(A) 35 layer ($\approx$6.4 nm) TlSbSe$_2$; (B) 47 layer ($\approx$ 8.6 nm) TlSbSe$_2$; (C)47 layer ($\approx$ 9.2 nm) TlSbTe$_2$; and 
(D)39 layer ($\approx$ 6.9 nm) TlSbS$_2$, along high symmetry lines in the surface Brillouin zone.  The bulk bands projected on the surface Brillouin zone are shown in blue color. }
\end{figure}

\subsection {TlBiQ$_2$ (Q=Se, Te and S)}
Similar to TlSbQ$_2$, TlBiQ$_2$ also shows thickness dependent slab electronic structure. Fig. 6 shows the surface band structure of TlBiSe$_2$. With a slab thickness of $\approx$6.3 nm we get a gapped surface state with a gap of 36 meV 
and as we increase slab thickness to $\approx$7.0 nm this gap reduces to 9 meV. The gap at small slab thickness arises due to interaction between the opposite faces of the slab. As we further
increase the slab thickness to $\approx$ 8.7 nm and $\approx$ 10.9 nm we obtain a clear Dirac state with a negligible gap. At a critical thickness of $\approx$ 7.0 nm the Dirac point is isolated,
i.e. no other states are at the energy of the Dirac point, which agrees with recent angle-resolved photo-emission spectroscopy (ARPES) \cite{chen}$^,$\cite{sato}$^,$\cite{kuroda} measurements.
The observed Dirac state for slabs thicker than 7.0 nm is surrounded by the surface bands, which is somewhat different from ARPES results and requires further study.
\begin{figure}[h!] 
\includegraphics[width=0.5\textwidth]{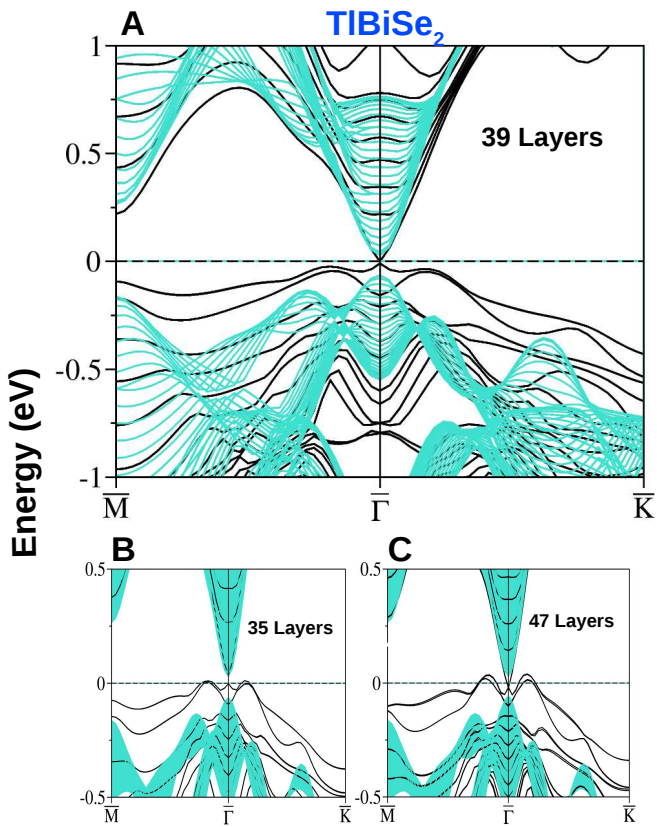}
\caption{Surface electronic structure of various slabs: (A) 39 layer ($\approx$ 7.0nm) TlBiSe$_2$; (B) 35 layer ($\approx$ 6.3nm) TlBiSe$_2$; and (C)47 layer ($\approx$ 8.7nm) TlBiSe$_2$, along 
high symmetry  lines in the surface Brillouin zone. Projected bulk bands are shown in blue color.} 
\end{figure}
\begin{figure}[h!]  
\includegraphics[width=0.5\textwidth]{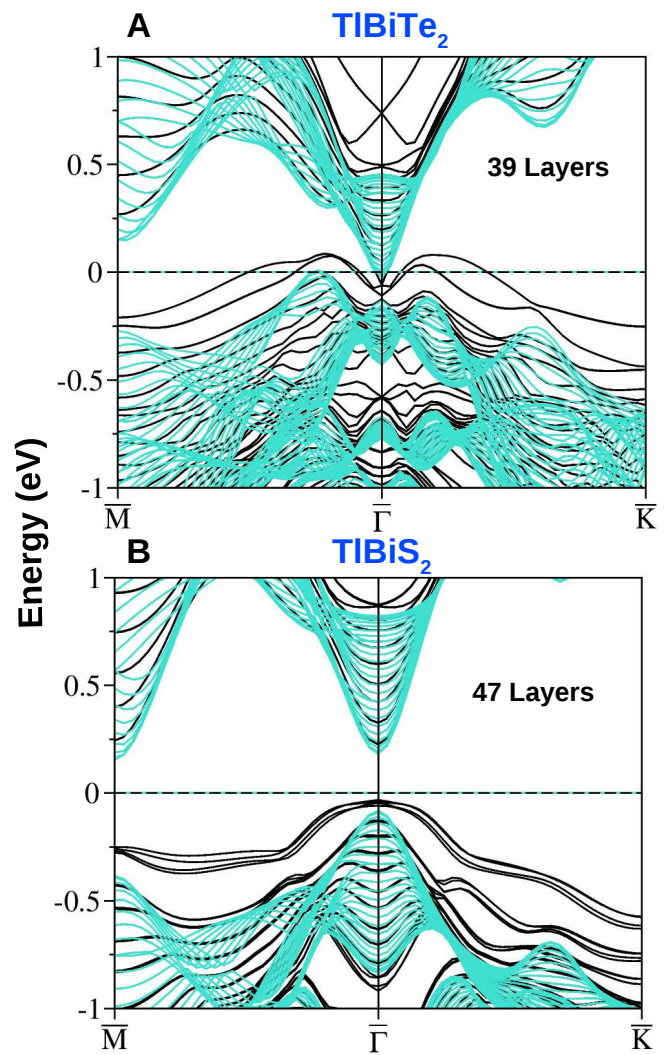}
\caption{Surface electronic structure of: (A) 39 layer ($\approx$ 7.4nm) TlBiTe$_2$ slab, and (B) 47 layer ($\approx$ 8.6nm) TlBiS$_2$ slab, along high symmetry lines in the surface Brillouin
 zone. The blue color shows projected bulk bands.}
\end{figure} 

Fig 7(A) shows results for TlBiTe$_2$, which exhibits a Dirac-like surface state in the bulk gap region at the thickness 
of $\approx$ 7.4 nm (39 layers). The Dirac point lies 0.1 eV below the Fermi energy in accord with ARPES studies \cite{chen}. The electronic structure 
displays an indirect band gap near the $\bar{\Gamma}$ point with bulk conduction band minimum at the $\bar\Gamma$-point and bulk valence band maxima along the
$\bar{\Gamma}$ $-$ $\bar{M}$ and $\bar{\Gamma}$ $-$ $\bar{K}$ directions.
The location of Dirac point in our calculation is close to experimental results, where Dirac point lies at 0.3 eV below the Fermi energy. Results for 
TlBiS$_2$ in Fig. 7(B) do not have any metallic surface state in the bulk
gap region indicating its topologically trivial character.

\begin{figure}[h!]  
\includegraphics[width=0.5\textwidth]{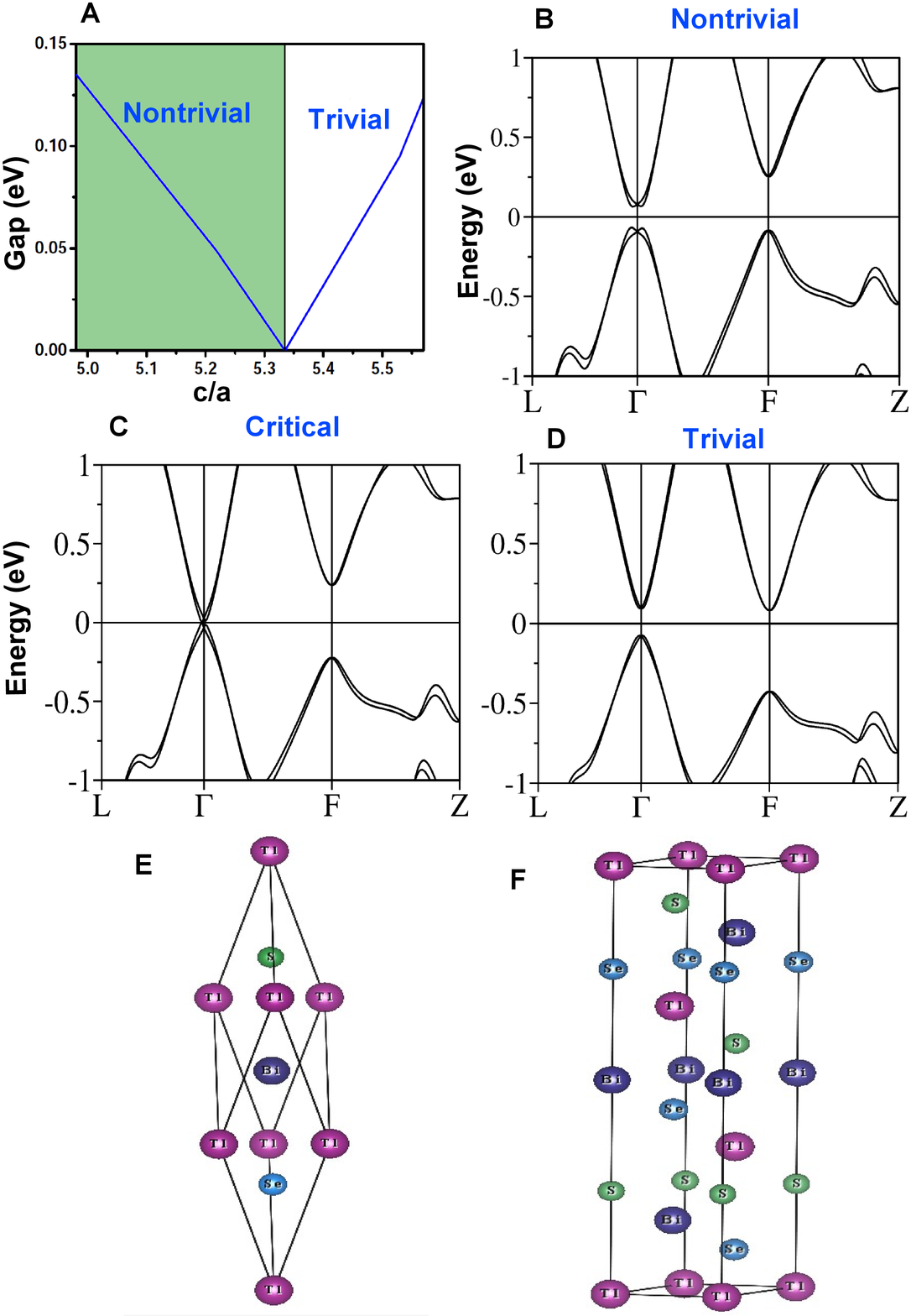}
\caption{ (A)Phase diagram for the topological phase transition. The system is tuned from the normal to topological insulator by changing the $c/a$ ratio. Panels (B)-(D) 
show the bulk band structure of TlBiSeS for $c/a$=5.22, $c/a$=5.33 and $c/a$=5.58, respectively. 
Primitive rhombohedral (E) and conventional hexagonal (F) crystal
structure for TlBiSeS. The layered structure is with no inversion symmetry.
}
\end{figure}

\vspace{.1in}
\section{Topological phase transition and Weyl semimetal }
\subsection{TlBi(S$_{1-x}$Se$_x$)$_2$}
The topological phase transition was observed in TlBi(S$_{1-x}$Se$_x$)$_2$ in which the S and Se atoms are disordered and inversion symmetry is preserved on the average with the inversion center at the (Se,S) site.\cite{mass} At the critical composition x$\sim$0.5, the conduction and valence bands meet at the $\Gamma$ point and form 3+1D Dirac-cone bulk states. The Dirac-cone states are doubly degenerate due to inversion and time-reversal symmetries. When the inversion symmetry is broken, the spin degeneracy can be lifted. Here we consider an ordered phase of TlBiSSe with layers in the order Tl-Se-Bi-S as shown in Fig.~8(E) and (F). The inversion symmetry is now seen to be broken as Se and S sites are no longer inversion centers.
Our theoretical fully relaxed structure, which is rhombohedral, is found to be stable, indicating that the material should be possible to realize via molecular beam epitaxy (MBE) techniques. In the absence of inversion symmetry, parity analysis cannot be applied to delineate the topological nature, but adiabatic continuity arguments can be used. Accordingly, we start from the 
normal insulator with a large value of $c/a$ and systematically reduce its value. During this process the system passes from normal to a topological insulator as shown in the Fig. 8(A). The band 
structure for the normal and topological phase through the critical point is shown in Figs. 8(B)-(D). The critical point occurs when the gap closes. At $c/a$=4.98, the band gap is 0.135 eV, and 
the bulk valence band has no $s$-character while the bulk conduction band possesses a finite $s$-character at $\Gamma$. These characteristic symmetries are the same as in TlBiSe$_2$, indicating 
that the two compounds are adiabatically connected and are thus both topologically nontrivial. With increasing $c/a$ value the gap decreases and becomes zero at the critical value $c/a$=5.33. 
Upon further increasing $c/a$, the bulk gap increases and becomes 0.129 eV at $c/a$=5.58 and the bulk valence and conduction bands swap their orbital character at the $\Gamma$-point. The system 
is now adiabatically connected to TlBiS$_2$ and is, therefore, topologically trivial.

Interestingly, at the topological critical point, the band gap at $\Gamma$ remains finite with a value of 0.062 eV. However, the band gap closes along  $\Gamma-L$ at $ \vec{k}$=(0.0, 0.0, 0.007).
Thus, by breaking inversion symmetry, we obtain a nondegenerate spin-polarized bulk Dirac cone, instead of the doubly degenerate Dirac cone found in the Se/S disordered TlBi(Se,S)$_2$.  The bulk
valence and conduction bands now touch each other and display linear dispersion, which can be described by a two component wave functions, i.e. by the Weyl equation. 
 A Weyl semimetal is formed at the critical point with six Weyl points centered along $\Gamma-L$.\cite{footWeyl}
The detailed band structure is shown in Fig. 9.

\begin{figure}[h!]  
\includegraphics[width=0.5\textwidth]{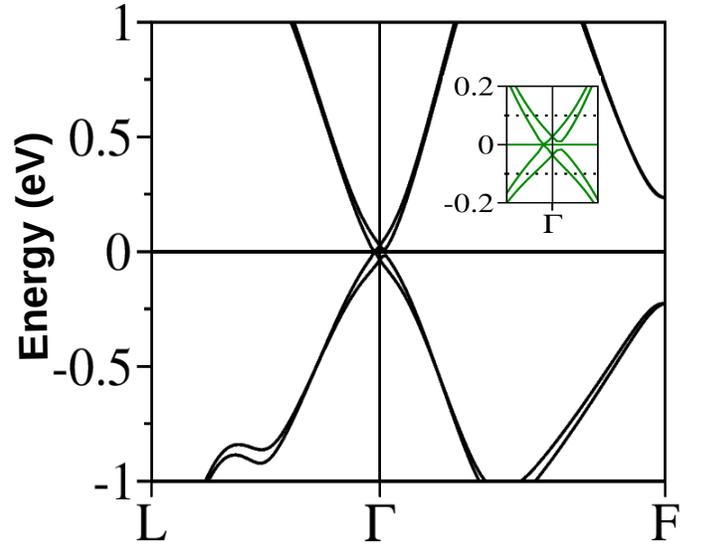}
\caption{Band dispersion at the critical point in TlBi(S$_{1-x}$Se$_x$)$_2$. Inset shows dispersion near the $\Gamma$-point. }
\end{figure} 

\subsection{TlBi(S$_{1-x}$Te$_x$)$_2$}
Analysis of TlBi(S$_{1-x}$Te$_x$)$_2$ follows along the lines of that of TlBi(S$_{1-x}$Se$_x$)$_2$ in the preceding subsection, and therefore, we only make a few relevant remarks.
 Here again we take the critical point to be at $x$=0.5, and break the inversion symmetry by 
 considering an ordered phase with layers in the sequence Tl-Te-Bi-S,  and compute the band structure for a series of 
$c/a$ values. The results, summarized in Fig. 10, show that at $c/a$=4.93 there is a direct band gap of 0.123 eV along $\Gamma-L$, although there is an indirect gap of $\approx 10 meV$.
 At the critical value of $c/a$=5.00, the gap becomes zero along $\Gamma-L$ at $\vec{k}=(0.0, 0.0, 0.025)$, even though the gap remain finite at $\Gamma$ with a value of 0.210 eV. With 
further increase in $c/a$, the gap reopens and becomes 0.221 eV at $c/a$=5.57. Adiabatic continuity arguments combined with the change in the orbital character of the bulk valence and 
conduction bands at the critical value of $c/a$ then allow us to conclude the non-trivial to trivial insulator transition shown in Fig. 10(A). As to the Weyl semimetal phase, at the 
critical $c/a$ value, TlBi(S$_{1-x}$Te$_x$)$_2$ displays six Weyl points along the $\Gamma-L$ direction at $\vec{k}=(0.0, 0.0, 0.025)$, although the spin splitting is larger than in
 TlBi(S$_{1-x}$Se$_x$)$_2$.

\begin{figure}[h!]  
\includegraphics[width=0.5\textwidth]{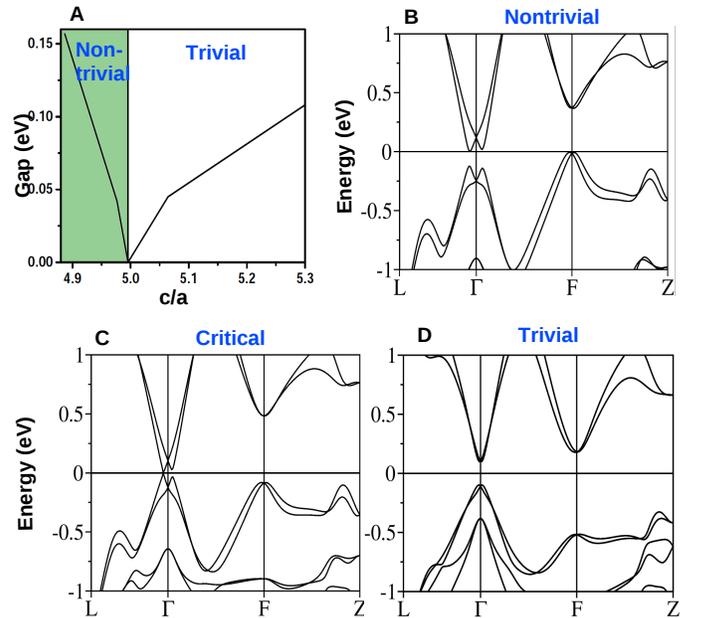}
\caption{(A) Phase diagram of topological phase transition in TlBiTeS. Panels (B)-(D) show the bulk band structure of TlBiTeS for $c/a$=4.93, $c/a$=5.00
and $c/a$=5.57, respectively}
\end{figure}

\section{Conclusions}
We have carried out an {\it ab initio} study of bulk and surface electronic structures of six thallium based III-V-VI$_2$ ternary chalcogenides TlMQ$_2$, where M(Bi, Sb) and 
Q(S, Se or Te), with focus on delineating the topological nature of these compounds. TlBiTe$_2$ is found to be a semi-metal with a band gap of -10 meV while the other five Tl-compounds are 
all small band gap semiconductors. Based on an analysis of parities of bulk band structures, we predict that TlSbSe$_2$, TlSbTe$_2$, TlBiSe$_2$ and TlBiTe$_2$ are non-trivial topological
insulators with band inversion at the $\Gamma$-point, but TlSbS$_2$ and TlBiS$_2$ are trivial band insulators. Moreover, surface state computations show that the surface Dirac states lie 
in the gap region at the $\Gamma$-point in all four aforementioned topological compounds. 
Our predicted topological phases and the $\Gamma$-point centered Dirac-cone surface states are in substantial accord with available ARPES results. 
Electronic structures of slabs with different numbers of layers were computed in order to gain insight into thickness-dependent effects. 
The gap opens at the Dirac point for thin slabs and decreases with increasing thickness. Finally, we investigated TlBi(S$_{1-x}$Se$_x$)$_2$ and TlBi(S$_{1-x}$Te$_x$)$_2$ alloys for $x$=0.5 where the inversion symmetry
was explicitly broken by using layers in the sequence Tl-Se(Te)-Bi-S and the $c/a$ ratio was varied. Both alloys were found to undergo a topological transition at a critical value of $c/a$ at 
which the spin degeneracy of the Dirac states is lifted and a Weyl semimetal phase could be realized with six Weyl points in the bulk Brillouin zone located along the $\Gamma-L$ directions. 

\section*{ACKNOWLEDGMENTS} 
We thank Diptiman Sen for helpful discussions. The work was supported by the Department of Science and Technology, New Delhi (India) through project SR/S2/CMP-0098/2010, 
the US Department of Energy, Office of Science, Basic Energy Sciences contract DE-FG02-07ER46352, and benefited from the allocation of supercomputer time at NERSC and Northeastern 
University's Advanced Scientific Computation Center (ASCC). MZH is supported by the Office of Basic Energy Sciences, U.S.  Department of Energy grant No. DE-FG-02-05ER46200 and the A. P. Sloan
Foundation Fellowship. 


\begin{thebibliography}{10}
\bibitem{footWeyl}
A topologically protected Weyl semimetal phase can be realized by removing the rotation symmetry of the crystal. The gapless semimetal phase is then stable over a finite region of a tuning parameter.
\end{thebibliography}

\end{document}